# Zipf Matrix Factorization : Matrix Factorization with Matthew Effect Reduction

Hao Wang
Ratidar.com
*Beijing, China*
haow85@live.com

*Abstract*—*Recommender system recommends interesting items to users based on users' past information history. Researchers have been paying attention to improvement of algorithmic performance such as MAE and precision@K. Major techniques such as matrix factorization and learning to rank are optimized based on such evaluation metrics. However, the intrinsic Matthew Effect problem poses great threat to the fairness of the recommender system, and the unfairness problem cannot be resolved by optimization of traditional metrics. In this paper, we propose a novel algorithm that incorporates Matthew Effect reduction with the matrix factorization framework. We demonstrate that our approach can boost the fairness of the algorithm and enhances performance evaluated by traditional metrics.*

## I. INTRODUCTION

Recommender system generates tremendous amounts of revenues for internet companies. Video apps such as TikTok and e-commerce websites like Amazon, among many other industrial firms, consider recommender system as their core technical assets. Recommender system predicts users' interest based on their past information history. By displaying such predictions to users to solicit clicks and purchases, websites decrease users' churn rate and increase users' retention rate.

Recommender system embodies a whole spectrum of techniques ranging from early days' collaborative filtering technologies to today's deep learning approaches. The algorithms aim to optimize evaluation metrics such as RMSE, MAE, AUC, nDCG and Precision@K. The efforts to increase algorithmic performance constitute the main evolutionary theme of recommender system technologies.

Matrix Factorization is a major technical breakthrough in the history of recommender system research. SVD [1], pLSA, LDA, ALS [2], SVD++ [3] and SVD Feature [4] are well-known matrix factorizations widely adopted in many industrial context settings. Matrix factorization techniques decompose the user-item rating matrix into the dot product of user feature vector and item feature vector. In this way, both the data complexity and computational complexity is reduced before the full user-item rating matrix is reconstructed for predictions. The computational complexity for an accurate matrix completion is high, but optimization techniques such as Stochastic Gradient Descent can increase the computation speed dramatically and are now standard approaches for solving the problem.

Although optimization of algorithmic performance has achieved fruitful research and practical outcomes, intrinsic problems of recommender systems such as Matthew Effect and fairness have largely been overlooked by the community until very lately. Like many social phenomena, input data of recommender system nearly always exhibit strong Matthew Effect problem. For example, at E-commerce website, popular products receive an astronomically larger number of clicks than unpopular products. When recommending products to e-commerce users, Matthew Effect of popularity of products can strongly contaminate algorithmic output.

In this paper, we propose a novel algorithm that incorporates the matrix factorization framework with Matthew Effect reduction. We show that by resorting to Matthew Effect reduction technique, the fairness of the algorithmic output result could be improved with increase of performance evaluated by other metrics such as MAE.

## II. RELATED WORK

Matrix Factorization is one of the most widely adopted recommender system algorithms. The algorithm decomposes the user-rating matrix into the dot product of the user feature vector and item feature vector. The loss function of matrix factorization techniques is universally RMSE. However, different approaches differ in the computation of the user and item feature vectors.

Conventional matrix factorization techniques include SVD, pLSA and LDA, among many others. There exist a whole spectrum of variants of these methods. Koren [5] incorporates time information into the matrix factorization framework and invents an algorithm known as timeSVD++ . A more holistic point-of-view of matrix factorization is invented by Chen [4] that solves the feature-based matrix factorization problem.

Research on Matthew Effect of recommender systems is recent. In SIGIR conferences ([6][7][8]), researchers start to model the Matthew Effect and fairness problem mainly in the framework of learning to rank. Himan et.al. [9] from University of Colorado at Boulder analyze these problems analytically. Wang et.al. [10] quantify the Matthew Effect and sparsity problem of collaborative filtering algorithms combinatorically and create a matrix algorithm MatRec [11] that incorporates Matthew Effect information into the problem formulation.

Matthew Effect as an important intrinsic problem of recommender systems has not attracted much attention from the community until very lately. Wang et. al. [10] computes combinatorial formulas that quantifies the Matthew Effect in user-based and item-based collaborative filtering algorithms. Canamares et. al. [12] utilizes probabilistic modeling techniques to model the Matthew Effect of Learning to Rank techniques. A matrix factorization approach consisting of Matthew Effect information is invented by Wang et. al. [11] with fairly accurate rating prediction and fast speed comparable with shallow models.

### III. PROBLEM FORMULATION

Most input data of recommender systems obeys power law distribution. The power law distribution is a common phenomenon existing in wealth distribution, social networks, machine learning algorithms' input structures, etc. To model the input data structure of recommender systems, we need first to model the power law effect of the input data structure. In order to model the power law effect, mathematicians invented a probabilistic distribution called Pareto Distribution. The distribution is framed as follows:

$$p(x) = \begin{cases} 0 & x < x_{\min} \\ \dfrac{k x_{\min}^k}{x^{k+1}} & x > x_{\min} \end{cases}$$

, where $x_{\min}$ and k are preset parameters of the distribution.

Pareto distribution is a continuous distribution for continuous variables. For discrete cases, linguists came up with a distribution called Zipf distribution. The Zipf distribution is derived from the observation that the $i^{th}$ most popular word in the English document corpus has a number of occurences proportional to $\dfrac{1}{i}$. The probability distribution of Zipf distribution is:

$$f(k;s,N) = \dfrac{1/k^s}{\sum_{n=1}^{N}(1/n^s)}$$

, where s and N are preset parameters of the distribution, k is the rank of the word. Zipf distribution is capable of capturing the disproportionally heavy head and long tail of the power law distribution, serving as an ideal choice for modeling the input data structure of recommender systems.

In a data distribution, we use the exponent s as the metric for evaluating the degree of Matthew Effect. The larger the value of s is , the stronger the Matthew Effect the data distribution exhibits. The aim of our algorithm is to maintain good performance of matrix factorization based on conventional evaluation metrics such as MAE and precision@K while reduce the Matthew Effect of output data structure, measured by the value of s.

Since the ouput data is not available during runtime of the algorithm, we need to estimate the Zipf Exponent s. We estimate the Zipf Exponent of output data structure as follows : we compute the distribution of item occurences F in the predicted user-item rating matrix of the matrix factorization algorithm. By Wang et.al.'s theory [10] , due to the Matthew Effect of the input data structure, the output data structure also exhibits Matthew Effect. A Zipf distribution is fit for F and the exponent is used as a penalty term for the loss function of matrix factorization algorithm. Newman [13] models the exponent of the Zipf distribution with a statistics, by which we do a small modification and define the statistics as follows (The rigid and detailed theory behind the statistic can be found in [14]):

$$s = 1 + n\left(\sum_{i=1}^{n} \ln \dfrac{x_i}{x_{\max}}\right)^{-1}$$

, where n is the number of items, $x_i$ is the popularity rank of the $i^{th}$ item, $x_{\max}$ is the largest rank of the items, namely, the number of items in the output data. The loss function of matrix factorization penalized by Zipf exponent is defined as :

$$L = \sum_{i=1}^{m}\sum_{j=1}^{n}\left(\dfrac{R_{ij}}{R_{\max}} - \dfrac{U_i^T \bullet V_j}{\|U_i\| \bullet \|V_j\|}\right)^2 - \beta\left(1 + n\left(\sum_{i=1}^{n}\ln\dfrac{x_i}{x_{\max}}\right)^{-1}\right)$$

, where $R_{ij}$ is the user-item rating score, $U_i$ is the user feature vector, $V_j$ is the item feature vector, $\beta$ is a penalty coefficient.

In this formulation, we need to approximate the rank of items in the output data structure, namely $x_i$ and $x_{\max}$. Since we do not know the output data beforehand, we need to use estimators to approximate the ranks. We approximate the rank of an item in the output data structure as the linear combination of user-item ratings involving this item :

$$x_j = \sum_{i=1}^{m} \alpha_i U_i^T \bullet V_j$$

However, the coefficients $\alpha_i$ of the linear combination are unknown in this equation as well. These coefficients are approximated as follows : We predict the user-rating matrix by the matrix factorization algorithm without Matthew Effect penalty and acquire the ranks of output items. We use these ranks to solve for coefficients of linear combinations using the same formula as the above one with a Lasso penalty term. After acquisition of the coefficients $\alpha_i$, we compute the values of $U_i$ and $V_j$.

We name the proposed matrix factorization after the name of Zipf distribution as Zipf Matrix Factorization. Stochastic Gradient Descent is used to solve for the optimal user feature vector and item feature vector with the formulas below:

$$U_i = U_i - \varepsilon \bullet \begin{pmatrix} \beta \bullet n/(\log(\alpha_{ij} \bullet t_0/n)^2 \bullet t_0) \bullet V_j \\ -2 \bullet (R_{ij} - t_0/t_3)/t_3 \bullet V_j \\ -2 \bullet t_0 \bullet (R_{ij} - t_0/t_3)/(t_1^3 \bullet t_2) \bullet U_i \end{pmatrix}$$

, where:

$$t_0 = U_i^T \bullet V_j$$
$$t_1 = \|U_i\|$$
$$t_2 = \|V_j\|$$
$$t_3 = t_1 \bullet t_2$$

, and :

$$V_j = V_j - \varepsilon \bullet \begin{pmatrix} \beta \bullet n/(\log(\alpha_{ij} \bullet t_0/n)^2 \bullet t_0) \bullet U_i \\ -t_4/t_3 \bullet U_i \\ +(t_0 \bullet t_4)/(t_2^3 \bullet t_1) \bullet V_j \end{pmatrix}$$

, where :

$$t_0 = U_i^T \bullet V_j$$
$$t_1 = \|U_i\|$$
$$t_2 = \|V_j\|$$
$$t_3 = t_1 \bullet t_2$$
$$t_4 = 2 \bullet (R_{ij} - t_0/t_3)$$

The optimization scheme of Zipf Matrix Factorization is simple and easy to implement. Any modern single or multi-core commercial machine could be used for the computation.

## IV. EXPERIMENTS

We compare Zipf Matrix Factorization algorithm with vanilla Matrix Factorization algorithm on MovieLens dataset with 610 users and 9742 movies.

When Zipf penalty coefficient is set to $10^{-3}$, Zipf Matrix Factorization achieves a best MAE score of 0.823113 when enumerating SGD learning rates while vanilla Matrix Factorization achieves a best MAE score of 0.843682. Zipf Matrix Factorization is superior to vanilla Matrix Factorization at its best performance. With different SGD learning rates, oftentimes Zipf Matrix Factorization achieves better MAE than vanilla Matrix Factorization Algorithm (Fig. 1).

When gradient learning step is set to $10^{-4}$, Zipf Matrix Factorization achieves the best MAE score of 0.8347304 when Zipf penalty coefficient is $7 \times 10^{-5}$. The MAE scores of Zipf Matrix Factorization drop below the best MAE score of vanilla Matrix Factorization multiple times with varying Zipf penalty coefficients (Fig. 2).

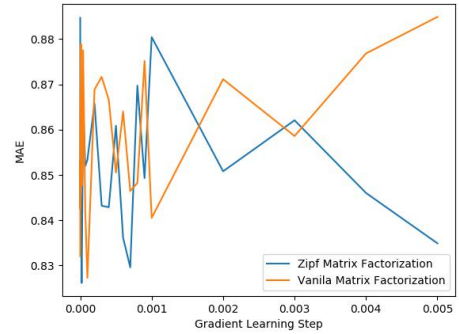

Fig.1 Zipf Matrix Factorization v.s vanilla Matrix Factorization with varying gradient learning steps

We compare the degree of Matthew Effect between two different methods using the following formula:

$$s = 1 + n(\sum_{i=1}^{n} \ln \frac{x_i}{x_{max}})^{-1}$$

The larger the value of s is, the weaker the Matthew Effect exhibits in the output data structure. When the gradient learning rate is fixed to $10^{-4}$, the Degree of Matthew Effect varies between -0.013 to -0.014 for Zipf Matrix Factorization, while the degree mostly drops below -0.08 for vanilla Matrix Factorization algorithm. Zipf Matrix Factorization yields better MAE performance while huge reduction the Matthew Effect in the output data structure (Fig. 3).

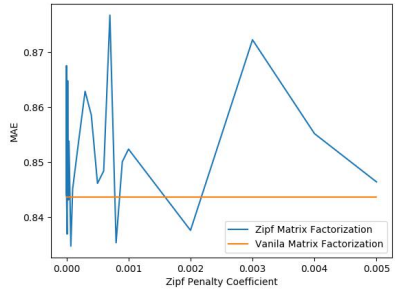

Fig.2 Zipf Matrix Factorization v.s best vanilla Matrix Factorization MAE score with varying Zipf Penalty Coefficient

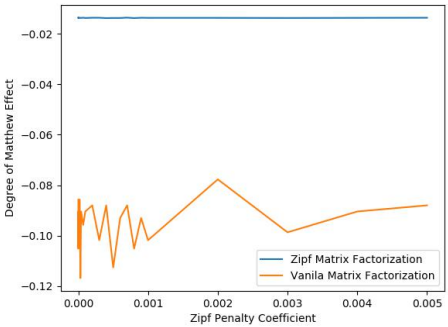

Fig.3 Zipf Matrix Factorization v.s Vanilla Matrix Fatorization on Degree of Matthew Effect

When industrial practitioners discuss social impact of recommender systems, the concern is usually the trade-off between social fairness and objective performance. Zipf Matrix Factorization is a reliever for the concern. The experimental results prove that Zipf Matrix Factorization not only can improve MAE performance of matrix factorization, but at the same time reduce the Matthew Effect of output data structures. The computational complexity of Zipf Matrix Factorization is on the scale of seconds on commercial laptops. The algorithm has great prospects of commercialization.

## V. CONCLUSION

Fairness problem in the appearance of Matthew Effect of recommender systems poses serious threats to successful materialization of online commercial systems. In this paper, we propose a matrix factorization method called Zipf Matrix Factorization that penalizes Matthew Effect in the loss function formulation.

In our experiments, we prove that the new method enhances the accuracy of score prediction of recommender systems while at the same time reduces the Matthew Effect of output data structures.

Obviously, the Zipf penalty can be generalized to other algorithms such as Factorization Machines, Learning to Rank and Deep Learning Approaches. We would like to explore the impact of Zipf penalty on such algorithms in our future work.

REFERENCES

[1] Y. Koren. 2009. The BellKor Solution to the Netflix Grand Prize.

[2] G. Takacs, D. Tikk. 2012. Alternating Least Squares for Personalized Ranking. The 6th ACM Conference on Recommender Systems.

[3] Y. Koren. 2008. Factorization Meets the Neighborhood: a Multifaceted Collaborative Filtering Model. KDD.

[4] T. Chen, W. Zhang, Q. Lu, K. Chen, Y. Yu. 2012. SVDFeature：A Toolkit for Feature-based Collaborative Filtering. The Journal of Machine Learning Research.

[5] Y. Koren. 2010. Collaborative filtering with temporal dynamics. Communications of ACM.

[6] M. Zehlike, C. Castillo. 2020. Reducing Disparate Exposure in Ranking: A Learning to Rank Approach. SIGIR.

[7] H. Yadav, Z. Du, T. Joachims. 2020. Fair Learning-to-Rank from Implicit Feedback. SIGIR.

[8] M. Morik, A. Singh, J. Hong, T. Joachims. 2020. Controlling Fairness and Bias in Dynamic Learning-to-Rank. SIGIR

[9] H. Abdollahpouri, M. Mansoury, R. Burke, B. Mobasher. 2020. The Connection Between Popularity Bias, Calibration, and Fairness in Recommendation. 14th ACM Conference on Recommender Systems.

[10] Hao Wang, Zonghu Wang, Weishi Zhang. 2018. Quantitative Analysis of Matthew Effect and Sparsity Problem of Recommender Systems, IEEE International Conference on Cloud Computing and Big Data Analysis,

[11] Hao Wang, Bing Ruan. 2020. MatRec: Matrix Factorization for Highly Skewed Dataset. The 3rd International Conference on Big Data Technologies.

[12] Rocío Cañamares, Pablo Castells. 2018. Should I Follow the Crowd? A Probabilistic Analysis of the Effectiveness of Popularity in Recommender Systems, SIGIR '18: The 41st International ACM SIGIR Conference on Research & Development in Information Retrieval,

[13] MEJ Newman. 2005. Power Laws, Pareto Distributions, and Zipf's Laws. Contemporary Physics.

[14] Albert-Laszlo Barabasi. 2016. Network Science. Cambridge University Press.